\documentclass[final,authoryear,5p]{elsarticle}

\usepackage{graphicx}
\usepackage{pdflscape}

\usepackage{amssymb}

\journal{Advances in Space Research}

\begin{document}

\begin{frontmatter}



\title{X-rays from magnetic intermediate mass Ap/Bp stars}

\author{Jan Robrade}
\address{Hamburger Sternwarte, Gojenbergsweg 112, 21029 Hamburg, Germany}
\ead{jrobrade@hs.uni-hamburg.de}

\begin{abstract}
The X-ray emission of magnetic intermediate mass Ap/Bp stars is reviewed and put into context of intrinsic as well as extrinsic hypotheses for its origin.
New X-ray observations of Ap/Bp stars are presented and combined with an updated analysis of the available datasets, providing the largest sample of its type that is currently available. 
In the studied stars the X-ray detections are found predominantly among the more massive, hotter and more luminous targets.
Their X-ray properties are quite diverse and beside strong soft X-ray emission significant magnetic activity is frequently present.
While a connection between more powerful winds and brighter X-ray emission is expected in intrinsic models, 
the scatter in X-ray luminosity at given bolometric luminosity is so far unexplained and
several observational features like X-ray light curves and flaring, luminosity distributions and spectral properties are often similar to those of low-mass stars.
It remains to be seen if these features can be fully reproduced by magnetospheres of intermediate mass stars.

The article discusses implications for magnetically confined wind-shock models (MCWS) and stellar magnetospheres under the assumption that the intrinsic model is applicable,
but also examines the role of possible companions. Further, related magnetospheric phenomena are presented and an outlook on future perspectives is given.
\end{abstract}

\begin{keyword}
X-rays: stars \sep Stars: Ap/Bp \sep Stars: intermediate mass
\end{keyword}

\end{frontmatter}

\parindent=0.5 cm

\section{Introduction}

The X-ray emission from stars can be roughly separated into two fundamental regimes.
Cool, low-mass stars of spectral type F to M have magnetic coronae that contain hot plasma at MK temperatures.
Magnetic activity is driven by stellar dynamos and related to convection and rotation and observed
activity levels cover a broad range of about $-7 < \log L_{\rm X}/L_{\rm bol} < -3 $.
The stellar dynamos become less efficient as the outer convective layer becomes shallower and magnetic activity steeply declines towards higher masses.
The X-ray emission of late A~stars is already quite faint and subsequently fades towards hotter stars.
For a discussion of coronal X-rays from the nearby A7 star Altair and similar objects, i.e. stars with masses of $\lesssim 2.0 M_{\odot}$, see e.g. \cite{rob09}.
In hot, massive stars of spectral type early B the instabilities in their radiatively driven winds become sufficiently strong to give rise to X-ray emission from wind shocks.
Basically, in O and early~B stars X-ray luminosity correlates with wind power and bolometric luminosity at a level of about $\log L_{\rm X}/L_{\rm bol} \approx -7$. 

Main-sequence stars of intermediate mass at spectral type mid A to mid B, also known as tepid stars, neither drive sufficiently strong winds to produce X-rays in wind shocks, nor possess an outer convection zone to generate magnetic activity and coronae via dynamo processes. Consequently they should be virtually X-ray dark and this is indeed likely true for all 'normal' main-sequence stars. See e.g. \cite{gued09} for a more detailed overview on stellar X-ray emission.

The Ap/Bp stars are magnetic intermediate mass stars that belong to the group of chemically peculiar (CP) stars.
The origin of their magnetic field is likely fossil, consistent with the finding that only a small fraction of about 5\,--\,10\,\% of the early-type stars are magnetic
A fossil origin of the magnetic field is supported by studies of pre-main sequence intermediate mass stars, the HAeBe stars, where the fraction of stars with large-scale magnetic fields is similar to those of main-sequence intermediate mass stars as shown by spectropolarimetric surveys \citep{ale13}.
In addition, the magnetic field strength of Ap/Bp stars and their rotation period are independent and the presence of abundance spots caused by diffusion indicates a stable photosphere,
both suggesting that the strong magnetism is not related to dynamo processes.
Furthermore, the magnetic fields in Ap/Bp stars are dominated by rather simple large scale structures like
a dipole and hence they are fundamentally different from the complex field geometry of magnetically active late-type stars, see e.g. the review by \cite{lan92}.

In Ap/Bp stars X-ray production may arise from magnetically confined wind-shocks (MCWS) as proposed by \cite{bab97} to 
describe the remarkably soft X-ray emission observed from the A0p star IQ~Aur. 
Already in the RASS (ROSAT All-Sky Survey) several Ap/Bp stars were detected as X-ray sources \citep{dra94}. 
The MCWS model was also successfully applied to magnetic massive stars like $\theta^{1}$~Ori~C
\citep{gag05}, but other studies ended up with more mixed results. 
In an X-ray study of magnetic late B- and A-type stars several objects remained undetected \citep{cze07}, showing that the pure existence of a strong magnetic field is not a sufficient criterion for X-ray emission.
Similarly, \cite{osk11} studied magnetic early B-stars and find that strong and hard X-ray emission not necessarily correlates with the presence of a magnetic field.

A systematic X-ray study of Ap/Bp stars by \cite{rob11} has addressed the puzzling mix of X-ray detections and non-detections among the Ap/Bp stars. 
Within the studied sample focussing on objects with $T_{\rm eff} \lesssim 15000$~K they find indications for a transition from X-ray dark (or faint) to X-ray emitting Ap/Bp stars
occurring at stellar luminosities of around 250~$L_{\odot}$. This temperature and luminosity regime corresponds roughly to the spectral boundary between B and A type stars, i.e. at significantly later spectral types than in non-magnetic objects. Further, in addition to wind-shocks also magnetic activity is clearly present in several of the studied objects. In the sense that an intrinsic
X-ray generation bases on magnetic fields and winds, Ap/Bp stars bridge the 'classical' X-ray regimes of cool and hot stars.

This review focusses on magnetic intermediate mass Ap/Bp stars of spectral types mid/late~B to early~A, in this respect complementing a study of magnetic massive stars by \cite{naze14}. 
The intermediate mass Ap/Bp stars share, if applicable, with the MCWS an X-ray generation mechanism with the magnetic massive stars (see review by ud-Doula \& Naz{\'e}, this volume), but provide
different environmental conditions with respect to magnetic field and wind strength. 
The paper is structured as follows. In Sect.~2 we introduce the MCWS model and discuss the prototypical Ap star IQ Aur in greater detail, 
in Sect.~3 we present the current observational status of X-ray emission from Ap/Bp stars as a class including an analysis of the available X-ray data.
It also covers a discussion of the X-ray emission in the context of current wind-shock models, the impact of possible low-mass companions and its relation to other magnetospheric phenomena. In Sect.~4 we give an outlook on future perspectives.

\section{The MCWS model in Ap/Bp stars}

An important milestone in the field of X-ray studies of Ap/Bp stars was a {\it ROSAT} PSPC observation that showed the A0p star IQ~Aur to be an X-ray source with an X-ray brightness of $L_{X}= 4 \times 10^{29}$~erg\,s$^{-1}$ \citep{bab97}. The derived plasma temperature of $T_{X} = 0.3$~keV is extraordinarily low for a star with this X-ray luminosity.
This combination of X-ray brightness and soft spectrum makes a possible low-mass companion, typically put forward as alternative explanation for unexpected X-ray detections of late B/early A stars, an unlikely explanation.
To explain the unusually soft X-ray spectrum of IQ~Aur they introduced the 'magnetically confined wind-shock' (MCWS) model, where
the radiatively driven wind components from both hemispheres are magnetically channelled and forced to collide in the vicinity of its equatorial plane. 
As a consequence, strong shocks of the nearly head-on wind collision produce sufficiently hot plasma and an equatorial disk is formed. Assuming an efficient conversion of kinetic energy this scenario can give rise to the observed X-ray emission under reasonable assumptions on the stellar wind parameters.
Basic predictions from this model are an X-ray luminosity of $L_{\rm X} = 1/2~\dot{M} V_{sh}^{2}$ and a plasma temperature of $T_{sh} = 1.13 \times 10^{5} [V_{sh}/(100~$km\,s$^{-1})]^{2}$ with $V_{sh}$ being the velocity at the shock (pre-shock), $\dot{M}$ the mass loss rate and $T_{sh}$ the temperature at the shock front (post-shock). 
Thus wind speeds of 400\,--\,600~km\,s$^{-1}$ generate X-ray emitting plasma with temperatures of 2\,--\,4~MK, at 950~km\,s$^{-1}$ plasma temperatures of 10~MK are reached
and a mass-loss rate of a few times $10^{-11} M_{\odot}$ is required to produce the X-ray emission of IQ~Aur.

This seminal idea motivated further developments and dynamic MHD versions of the MCWS model were created and consecutively extended \citep{dou02, dou08, dou14}.
Up to date versions include e.g. stellar rotation or plasma cooling effects. Adding more physical details also revealed new phenomena and lead to a much more diverse picture than the simple scaling
laws of the original model. For example the so-called 'shock retreat', an effect that is especially important for weaker winds, can modify the X-ray luminosity and might lead in its extreme
to a total quenching of X-ray emission. As a consequence, the models by \cite{dou14} require typically a factor of a few up to an order of magnitude higher mass loss rates compared to the models from \cite{bab97}, to obtain a similar X-ray luminosity.
On the other hand, all magnetospheric models show that for strong magnetic fields naturally rigidly rotating disk like structures are formed in the magnetosphere of these stars.
The presence of a disk harboring centrifugal magnetosphere provides an environment for a variety of phenomena in addition to
wind-shocks, among these are rotational modulated emission, magnetic activity or flares.

The importance of magnetic field effects on the stellar wind is described by a characteristic 'wind magnetic confinement parameter' $\eta_{*}$, given as the ratio between the energy density of the magnetic field and the kinetic wind energy density \citep{dou02}. It is defined as $\eta_{*} = B^{2}_{\rm eq}R^{2}_{*}/\dot{M}V_{\infty}$, where $B_{\rm eq}$ is the equatorial field strength, $R_{*}$ the stellar radius, $\dot{M}$ the mass loss rate and $V_{\infty}$ the terminal wind speed. 
For strongly magnetic objects the rigid-field hydrodynamics (RFHD) approach provides an alternative to full MHD magnetospheric models \citep{tow07}.
Note however, that present simulations are typically tailored to stars more massive than the ones discussed here. 
Magnetospheres of rotating stars can also be characterized by a comparison of the respective Kepler co-rotation radius ($R_{K}$) and the Alf{\'v}en radius ($R_{A}$).
Assuming a bipolar field and aligned rotation, one finds that
within $R_{K}$ gravity always dominates and matter falls back onto the star and beyond $R_{A}$ the wind dominates, opens the field lines and matter escapes outwards.
However, if $R_{A}> R_{K}$ is fulfilled, in between these two radii matter can be trapped in closed magnetic structures and accumulate around the stellar magnetic equator in a disk like structure. A classification based on this criteria into centrifugal magnetosphere (CM, $R_{A}> R_{K}$) and
dynamical magnetosphere (DM, $R_{A}< R_{K}$) provides a useful separation that characterizes many key observables of massive stars \citep{pet13}.

The studied Ap/Bp stars have strong magnetic fields and weak winds.
The condition of strong magnetic confinement ($\eta_{*} \gg 1$) is always fulfilled, typical values are $\eta_{*} \sim 10^{6}$, and they reside in the 'CM regime'.
Disk harboring centrifugal magnetospheres have also been evoked to explain dynamic phenomena like magnetic reconnection and X-ray flares \citep{dou06}.
For example an ongoing debate exists about the X-ray emission and repeated flaring seen from the B2p star $\sigma$\,Ori~E. While \cite{san04} proposed a companion as flare location,
and $\sigma$\,Ori~E has a known companion, \cite{mul09} argue for an origin of the flare on the Bp star itself. 

The actual mass-loss rates of Ap/Bp stars are very uncertain and especially for weaker winds hardly accessible via direct measurements and the properties of magnetospheric disks are poorly known.
It remains to be shown that an extrapolation of the current MCWS models to intermediate mass stars like IQ~Aur is valid and to what extend the specific dynamic
magnetospheres will contribute to the X-ray properties. If applicable,
a combination of strong shocks in the wind collisions and the presence of magnetically confined plasma in a dynamic circumstellar disk 
might explain the diverse X-ray properties as observed in different Ap/Bp stars. The details for each individual object then
depend on the respective stellar parameters, magnetic field properties and viewing geometry.

\begin{figure}
\begin{center}
\includegraphics[width=84mm]{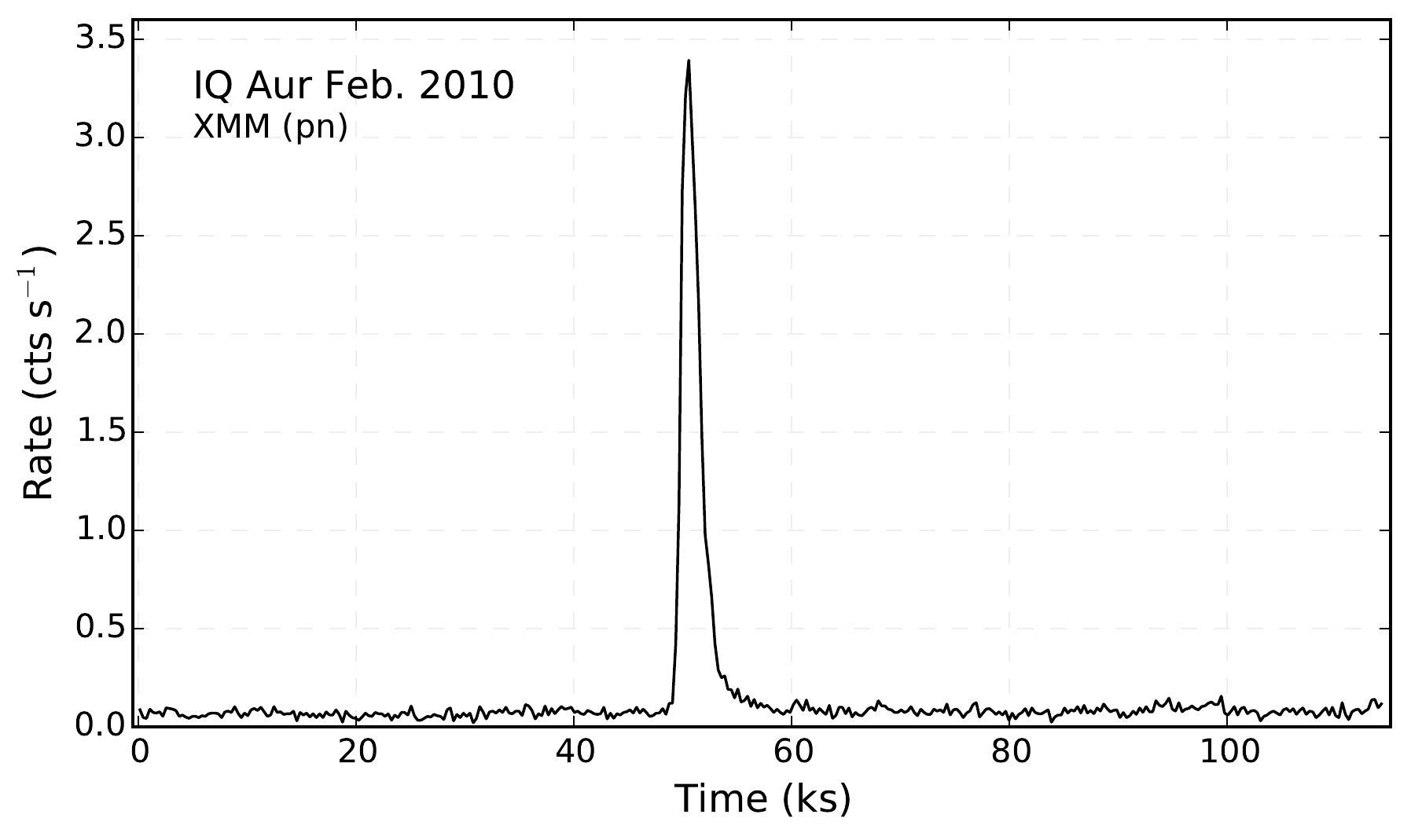}
\end{center}
\caption{\label{figiqaur}X-ray light curve of IQ~Aur; XMM-Newton pn data with 300~s binning. See \cite{rob11} for a detailed analysis of the IQ Aur data.}
\end{figure}

\subsection{The case of IQ Aur}

A deep XMM-Newton observation confirmed IQ~Aur as a bright X-ray source with $L_{\rm X} = 5 \times 29$~erg\,s$^{-1}$ 
and allowed to constrain its X-ray properties in great detail \citep{rob11}. It is the best studied Ap star at X-ray energies and in addition a sufficient high-resolution
X-ray spectrum is available. An analysis of emission line ratios like OVIII/OVII shows a pronounced excess of soft emission compared to coronal sources, 
confirming the exceptional X-ray properties of IQ~Aur. However,
beside signatures that match quite well with the standard MCWS mechanism, clear signs of magnetic activity are seen in these data. This includes a strong X-ray flare (see Fig.~\ref{figiqaur}) with 
an X-ray flux increase by about two orders of magnitude that is accompanied by the occurrence of very hot plasma with temperatures up to 100~MK during the maximum of the event.
This behavior is hardly explainable by wind-shocks alone and clearly points to ongoing magnetic activity.
In the context of the MCWS model flares would be associated to centrifugal break-out event that disrupt parts of 
the circumstellar disk or to reconnection in the circumstellar magnetosphere. In addition, the obtained X-ray grating spectrum revealed that the detected
X-ray emission originates predominantly from regions at least several stellar radii above the surface of IQ~Aur, 
fully consistent with the location of the predicted magnetospheric disk structures and wind shock zones at several stellar radii away from the star.

IQ~Aur is thought to be a single star, no visual companion is known and radial velocity searches with a sensitivity of a few km\,s$^{-1}$ by \cite{abt73} resulted in a constant velocity.
Nevertheless, the large flare motivated us to check for a possible late-type companion with adaptive optics imaging.
We initiated a dedicated companion search with MMT-AO/ARIES, but also failed to detect any companion. 
Still, a low-mass companion to IQ~Aur might just have escaped detection so far. 
The obtained brightness limits are $\Delta K \sim 3\,/\,6$ at distances above $0.3\,^{\prime\prime}$/$0.9\,^{\prime\prime}$ separation.

While the observed X-ray luminosity could be generated by a hypothetical companion, an active mid/late K dwarf would be sufficient, some problems exists with this interpretation.
When assigning the observed X-ray properties of IQ~Aur to a low-mass stars, one would also require an extraordinary
companion that differs from other known coronal sources. This does not favor the exterior hypothesis as a sole explanation.
Also the IQ~Aur flare energetics are extreme and even suitable active stars produce less than one of these events per year, challenging also the flare origin on a low-mass star.
Finally, it is also not ruled out that two components contribute to the X-ray emission; a low-mass companion might produce predominantly the strong magnetic activity 
and the magnetosphere adds predominantly to the softer X-ray emission.

\subsection{The Ap/Bp star puzzle}

IQ~Aur belongs to the class of $\alpha^2$\,CVn rotationally variable stars, but
the prototypical star $\alpha^2$\,CVn turned out to be X-ray dark \citep{rob11}.
Since both are optically very similar stars of spectral type A0p one might expect also rather similar X-ray properties, however their X-ray luminosities differ by at least three orders of magnitudes. 
Also several other attempts to detect X-rays from Ap/Bp stars failed or succeeded in an apparently random manner, raising the question if the
X-ray emission originates from companions or if MCWS X-rays are an atypical or transient phenomenon.
Further, individual objects might not be representative for the stellar class as the influence of the various stellar characteristics on the final X-ray properties of a particular star is until now only poorly understood.

After it became clear that spectral classification or strong magnetic fields are no sufficient criteria for X-ray emission,
a more systematic investigation of Ap/Bp stars at X-ray energies was required to make any progress. To study more general trends,
X-ray observations of similar intermediate mass magnetic stars were collected and combined to a sample of Ap/Bp stars as presented in \cite{rob11}.
This first sample allowed to study tentative trends, that relate X-ray emission to stellar parameters like bolometric luminosity.

\section{X-ray observations of Ap/Bp stars}

Building upon our initial study, we have extended the Ap/Bp stars sample with new datasets that are presented here for the first time.
This study uses X-ray observations of Ap/Bp stars that were carried out with Chandra and XMM-Newton and an updated analysis is performed for all targets.
Several X-ray observing campaigns of selected magnetic intermediate mass stars were conducted out within the last years and
combined with available data in the archives the full sample of X-ray studied Ap/Bp stars is now of moderate size.
This analysis focusses on stars with spectral types in the range mid/late-B to early-A, i.e. masses of about 2.5\,--\,6.0~$M_{\odot}$ or likewise effective temperatures of 10\,--\,20~kK and luminosities of  80\,--\,1000~$L_{\odot}$.
This selection leads to a stellar sample with about 30 observed objects (see Table~\ref{obs}). Data processing uses the standard software packages Chandra/CIAO and XMM/SAS and the performed analysis is analog to the procedures described in \cite{rob11}. A description of the instruments, software and data archives can be found 
under {\it http://cxc.harvard.edu} (Chandra) and {\it http://xmm.esac.esa.int} (XMM).

For our sample stars X-ray properties are obtained from modelling their medium resolution CCD-spectra to characterize their X-ray emission.
All stars with a sufficient number of detected X-ray photons are modeled with XSPEC and we use single or multi-temperature APEC plasma models to derive X-ray fluxes, temperatures and emission measures.
At very few detected counts, the obtained results were cross-checked with count-rate to flux conversion using the WebPIMMS tool. 
Several of the presented dataset have been analyzed before; beside in \cite{rob11} an analysis of X-ray data from some of the here discussed stars
can also be found in \cite{ste06} and \cite{cze07}.
At the hot end (mid B stars) the presented stellar sample overlaps with the one of the massive-stars magnetosphere program as discussed in  \cite{pet13,naze14}.

In the following comparative analysis the focus is on Ap/Bp stars with overall well determined properties.
We adopt stellar parameters from \cite{koch06} if available; additional data is taken from \cite{lan07}. This selection procedure slightly reduces the number of stars (24/28), 
but ensures a more consistent dataset. Due to their nature as CP stars, the modeling of Ap/Bp stars is a rather complex affair in contrast to their non-magnetic siblings and determined stellar parameter are by far not unique in literature.
The spectral/CP class and magnetic field data (average magnetic field) are from \cite{by09}, alternative spectral/CP classifications are from \cite{ren09}.
Some distances used in these publications have been slightly revised by a re-analysis of Hipparcos data, but typically deviations are a few percent and
for consistency we use the same distances as those in calculating the stellar parameters.

Stellar parameters and results from modelling of the X-ray data are summarized in Table\,\ref{results}. 
All given X-ray properties refer to the 0.2\,--\,5.0~keV band if not stated otherwise, $kT$ denote the emission measure weighted average temperature if multi temperature models were applied.
As an indication of the respective data quality we quote the approximate number of detected X-ray photons with XMM/pn or Chandra/ACIS detector.

\begin{table}[t]
\caption{\label{obs}X-ray observations of Ap/Bp stars.}
\setlength\tabcolsep{5pt}
\begin{tabular}{llrrr}
\\\hline
Star   & Observatory & Obs-ID & Year & $t$\,(ks)\\\hline
HD 12767  & Chandra & 5391   & 2005 &  3 \\ 
HD 22470  & Chandra & 13615  & 2011 & 11 \\
HD 25823  & Chandra & 13614  & 2012 & 12 \\
HD 27309  & XMM & 0201360201 & 2004 & 43 \\
HD 28843  & Chandra & 13616  & 2012 & 11 \\
HD 34452  & XMM & 0600320101 & 2010 &116 \\ 
HD 40312  & Chandra & 13617  & 2011 & 10 \\ 
HD 54118  & XMM & 0655220201 & 2010 & 21 \\
HD 73340  & Chandra & 4492   & 2004 &  3 \\
HD 75049  & XMM & 0651230201 & 2010 & 25 \\
HD 92664  & XMM & 0740290301 & 2014 & 36 \\
HD 105382 & XMM & 0742340101 & 2014 & 10 \\
HD 110073 & Chandra & 3741   & 2003 &  3 \\
HD 112413 & Chandra & 9923   & 2009 & 15 \\
HD 124224 & XMM & 0677980501 & 2011 & 28 \\
HD 125823 & Chandra & 13618  & 2012 & 10 \\
HD 133880 & Chandra & 2543   & 2002 &  2 \\
HD 137509 & XMM & 0651230101 & 2010 & 16 \\
HD 142301 & XMM & 0142630301 & 2003 & 22 \\
HD 143473 & Chandra & 5385   & 2005 & 15 \\
HD 144334 & Chandra & 5393   & 2005 &  3 \\
HD 146001 & Chandra & 5394   & 2005 &  3 \\
HD 147010 & Chandra & 5386   & 2006 & 15 \\
HD 175362 & Chandra & 13619  & 2012 & 11 \\
HD 184905 & Chandra &  5387  & 2005 & 15 \\ 
HD 208095 & Chandra &  5388  & 2005 & 15 \\
HD 215441 & Chandra &  5389  & 2005 & 20 \\
HD 217833 & Chandra &  5390  & 2006 & 15 \\\hline
\end{tabular}
{\small All Chandra observations use the ACIS detector without any grating inserted. 
XMM/pn data were mostly obtained with the thick filter, a few optically fainter targets use the medium filter.
Instrument descriptions and data archives at: {\it http://cxc.harvard.edu} and {\it http://xmm.esac.esa.int}.}
\end{table}

Binarity is a possible issue and we checked the targets for multiplicity by using the WDS (Mason+ 2001-2014, \cite{mas01} for visual multiples and SB9 (Pourbaix+ 2004-2014, \cite{pou04} as well as (Chini+ 2012, \cite{chi12} for spectroscopic double or multiple stars. 
Stars classified as spectroscopic binaries of type SB2 are likely binaries with components of similar mass and since non-magnetic late B to mid A stars are virtually X-ray dark they unlikely contribute to the X-ray emission. Note that also some of the undetected systems are multiples of type SB2, strengthening the conclusion
that roughly equal mass companions do not play a major role in our study.
If an object is characterized as SB1 or as multiple system with a suspected late-type companion, these could play a significant role in shaping the X-ray properties of the system.
As noted by \cite{abt73}, Ap stars have a significantly low spectroscopic binary fraction and the formation of close binaries magnetic stars is strongly suppressed in magnetic stars.
The visual binary fraction appears to be normal and
in several cases visual binary systems are resolved in X-rays and the analysis shows that indeed late-type companions can dominate the X-ray flux of the system.

\begin{figure}
\begin{center}
\includegraphics[width=82mm]{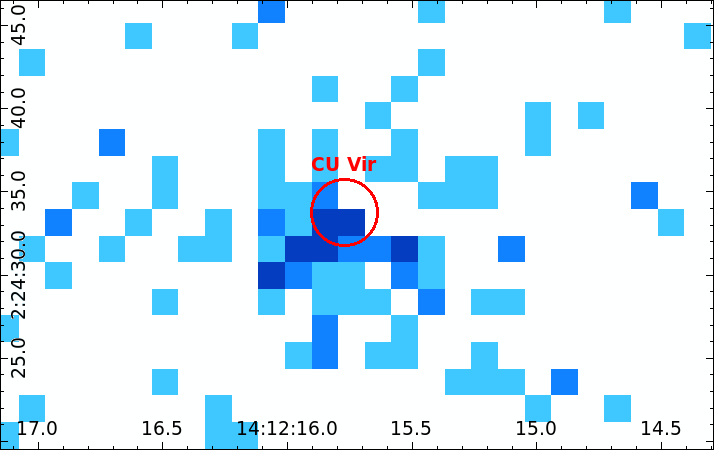}
\end{center}
\caption{\label{cuvir}MOS1 image of CU~Vir, the optical position is indicated by the circle with 2$^{\prime\prime}$ radius.}
\end{figure}

In the following we give some
notes on individual stars that are either not included in Table~\ref{results} due to very uncertain stellar parameters, where the X-ray detection is marginal/dubious 
or which are known multiple systems.

\noindent 
HD 12767 ($\nu$ For), SB2 (Chini+, 2012).
Detected, contamination from secondary possible.
 
\noindent 
HD 25823 (GS Tau), SB1 (SB9 Pourbaix+, 2004-2014).
Detected, contamination from secondary possible.
 
\noindent 
HD 40312 ($\theta$ Aur), binary (A0p + G2, sep. $2.1\,^{\prime\prime}$) in WDS.
Fully resolved, only secondary detected with $L_{\rm X} = 2 \times 10^{29}$~erg\,s$^{-1}$ at 51~pc.
 
\noindent 
HD 73340 (HV Vel), close binary (B8 Si + F?, sep. $0.6\,^{\prime\prime}$) in WDS.
Detected with $L_{\rm X} = 2 \times 10^{30}$~erg\,s$^{-1}$ at 140~pc.
Not fully resolved, X-ray position suggests secondary as main/sole contributor.

\noindent 
HD 75049, Ap star with extreme magnetic field of 30~kG \citep{fre08}, close K-type binary in WDS.
Detected with $L_{\rm X} = 1 \times 10^{29}$~erg\,s$^{-1}$ at 350~pc, distance and stellar parameters very uncertain/unknown.
 
\noindent 
HD 105382 (HR 4618), SB2 (Chini+, 2012).
Detected, contamination from secondary possible.
 
\noindent 
HD 109026 (gam Mus), SB2 (Chini+ 2012), (B5 + Ap?).
Not detected, B5 He-w is misidentification, primary is not magnetic \citep{ale14}.
 
\noindent 
HD 110073 (I Cen), multiple (B8p + F?, sep. $1.2\,^{\prime\prime}$) in WDS.
Detected, primary with $L_{\rm X} = 9 \times 10^{28}$~erg\,s$^{-1}$ at 140~pc, stellar parameters uncertain.
 
\noindent 
HD 112413 ($\alpha^{2}$ CVn), wide double (A0p + F0, sep. $19.2\,^{\prime\prime}$).
Fully resolved, only secondary detected with $L_{\rm X} = 4 \times 10^{28}$~erg\,s$^{-1}$ at 34~pc.
 
\noindent 
HD 124224 (CU Vir), B8/A0p star.
Detected with $L_{\rm X} = 2 \times 10^{28}$~erg\,s$^{-1}$ at 80~pc and $T_{\rm X} \sim 1.3$ keV). The source is slightly offset, but deviations are within limits;
looks a bit fuzzy and might be multiple (see Fig.\,\ref{cuvir}). With caution we attribute the X-ray detection to CU~Vir. 
Stellar radio pulsar and very fast rotator \citep{tri11}, quite broad range of stellar parameters in literature.
 
\noindent 
HD 133880 (HR Lup), binary (B8+ F?, sep. $1.2\,^{\prime\prime}$) in WDS.
Detected  with $L_{\rm X} = 1 \times 10^{30}$~erg\,s$^{-1}$ at 126~pc.
Not fully resolved, X-ray position suggests secondary as main/sole contributor (also \cite{ste06}).

\noindent 
HD 137509 (NN Aps), B9p with extreme and complex magnetic field of 29~kG \citep{koch06a}. X-ray detection marginal with weak excess in softer X-ray band only.
UL (detection)  with $L_{\rm X} = 3 (1) \times 10^{28}$~erg\,s$^{-1}$ at 250~pc.
 
\noindent 
HD 208095 (HR 8357), SB9 + wide binary B6 + A1p (HD 208063) in WDS.
All components undetected, distance and stellar parameters very uncertain.
 
\noindent 
HD 215441 (GL Lac, Babcock's star), B9/A0p star with extreme magnetic field of 34~kG \citep{bab60}.
Detected with $L_{\rm X} = 2 \times 10^{30}$~erg\,s$^{-1}$ at 1~kpc, distance and stellar parameters very uncertain.
 
\noindent 
HD 217833 (V638 Cas), close binary (B9 Si + F?, sep. $0.6\,^{\prime\prime}$) in WDS.
Detected with $L_{\rm X} = 7 \times 10^{28}$~erg\,s$^{-1}$ at 160~pc.
Not fully resolved, X-ray position suggests secondary as main/sole contributor (also \cite{cze07}).

\subsection{Global X-ray properties}

\begin{figure*}[t]
\begin{center}
\includegraphics[width=92mm]{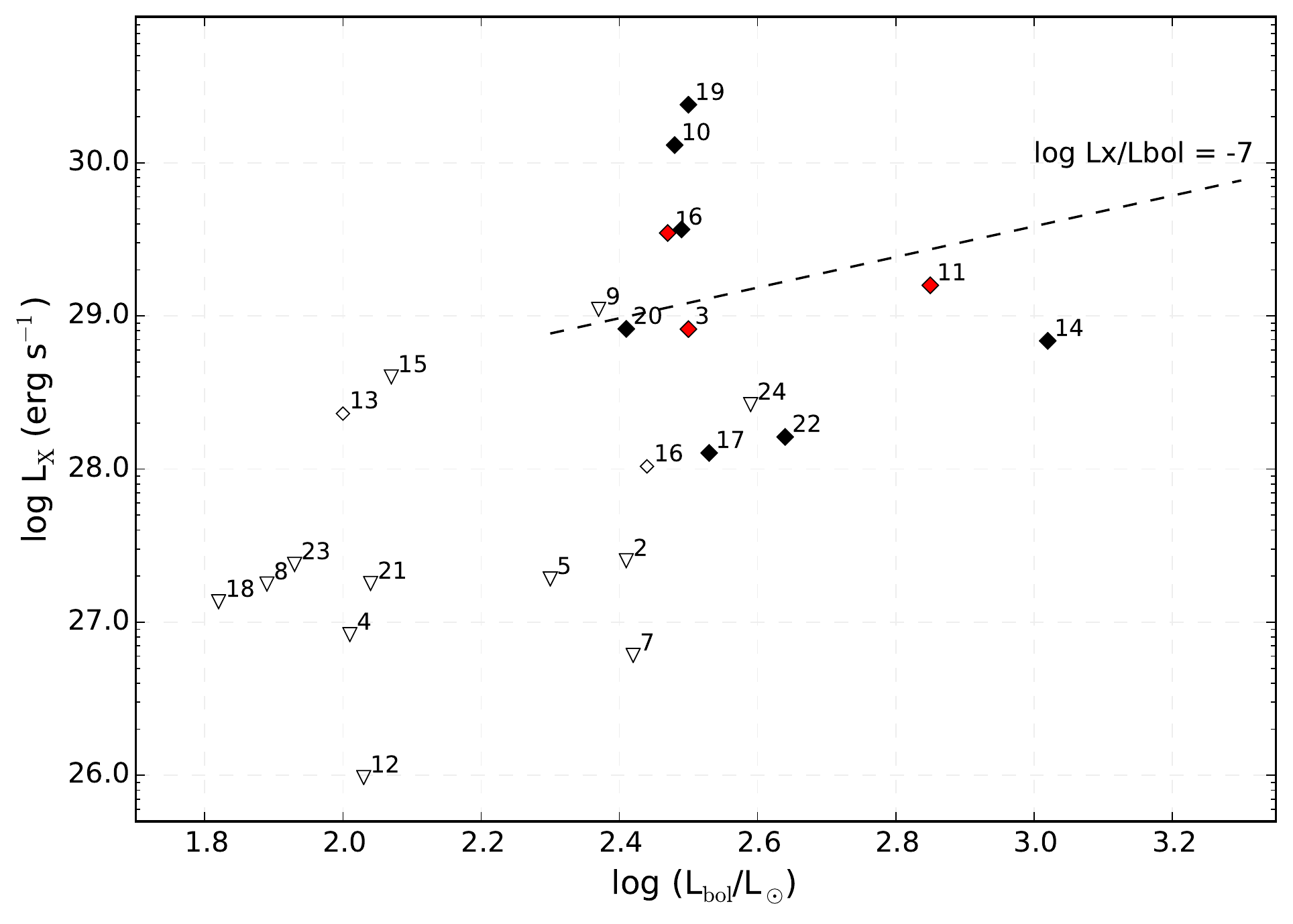}
\includegraphics[width=90mm]{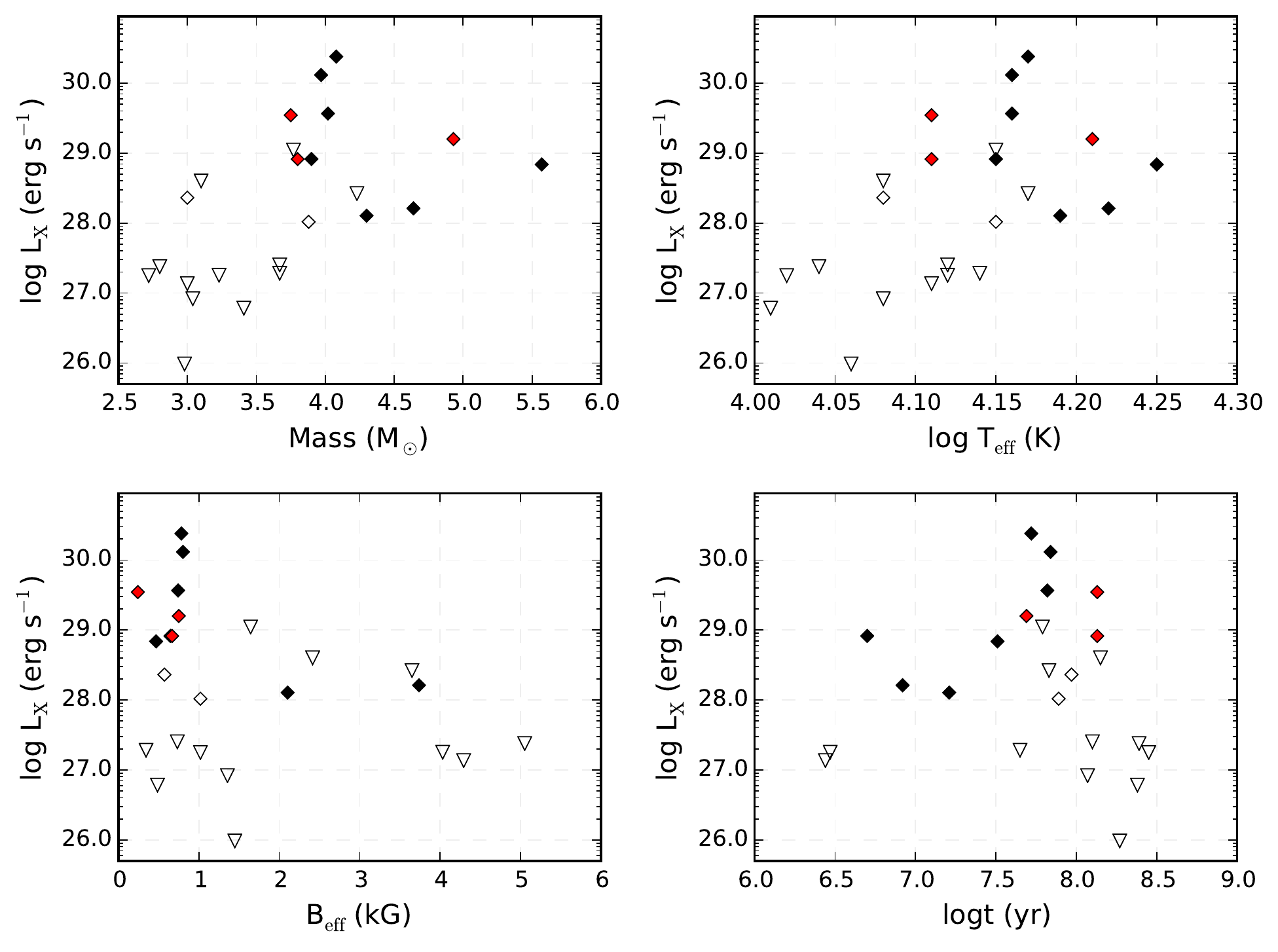}
\end{center}
\caption{\label{figlx}{\it Left:} X-ray luminosity vs. bolometric luminosity of Ap/Bp stars, diamonds denote detections (black: likely single, red: spectroscopic binary, open: marginal/ambiguous) and upper limits as open triangles. The number attached to each data point denotes the ID from Table~\ref{results}.
{\it Right:} X-ray luminosity vs. stellar parameters (mass, eff. temperature, eff. field strength, age), symbols as in left panel. See text for details.}
\end{figure*}

The X-ray brightness of the sample Ap/Bp stars compared to their optical luminosity is shown in the left panel in Fig.~\ref{figlx} and
it is striking that the X-ray detected stars are found virtually exclusively among the brighter targets above 250 solar luminosities.
In the bright fraction of the sample there are predominantly X-ray detections, while in the faint one there are predominantly X-ray upper limits, likely reflecting a predominantly
X-ray active and an X-ray faint (or even dark) stellar population. Inspecting the distribution in more detail, one
finds that all stars with $\log L_{\rm bol} \gtrsim 2.5~L_{\odot}$ are detected in X-rays
at $\log L_{\rm X} \gtrsim 28.0$~erg\,s$^{-1}$ and that the upper limits for these stars are above the fainter detections.
Stars with $\log L_{\rm bol} \lesssim 2.3~L_{\odot}$ are typically X-ray dark with upper limits up to one to two orders of magnitudes below the faintest detections.
In the intermediate luminosity range detections and non-detections are mixed and the detected object have a broad range of X-ray luminosities from $\log L_{\rm X} \approx 28.0$~erg\,s$^{-1}$ up to  $\log L_{\rm X} > 30.0$~erg\,s$^{-1}$.

There is some recurrent debate, if the X-ray emission from these objects is intrinsic or extrinsic.
While companions cannot be ruled out for any of the detected targets on an individual basis,
one can utilize statistical arguments to address this question with the existing data sample.
Neglecting upper limits above $\log L_{\rm X} = 28.0$~erg\,s$^{-1}$, i.e the least sensitive observations, we find above $\log L_{\rm bol} = 2.45~L_{\odot}$ a detection rate of 100\,\% (9/9)
while below $\log L_{\rm bol} = 2.35~L_{\odot}$ it is only 10\,\% (1/8). Assuming that the stellar sample is representative,
the probability that this outcome is by coincidence is extremely slim. Also it cannot be mimicked by evolving companions.
While magnetic activity declines with age and the less massive Ap/Bp stars have longer main-sequence lifetimes,
at the here relevant ages  of 3\,--\,300~Myr virtually all low-mass stars emit X-ray at a level well above $\log L_{\rm X} = 28.0$~erg\,s$^{-1}$ \citep[e.g.][]{gar11}.
If the sample is not severely biased by unknown reasons, the statistical argument supports an intrinsic X-ray generating mechanism in favor of the extrinsic hypothesis,
a finding should at least hold for the global observed trend.
Adopting the intrinsic model for the moment, this then suggests that stellar parameters of the Ap/Bp stars lead to a transition from an X-ray emitting to a virtually X-ray dark population.
It is obvious, that at some point when mass-loss rates become too low and/or the winds becomes too slow, a break-down of observable X-ray emission has to occur in any of the MCWS models.
The current data then suggests, that this occurs around spectral types late~B/early~A.

Although a prime dependence on stellar luminosity is plausible, the apparent transition is quite abrupt and a large scatter in X-ray luminosity among the detected stars is present.
As a guidance the 'classical' wind X-ray line of single, non-magnetic hot stars at around $\log L_{\rm X}/L_{\rm bol} = -7$ is plotted in Fig.~\ref{results}.
A few stars align around this level while a others are around $\log L_{\rm X}/L_{\rm bol} = -8$, but also at higher X-ray emission levels are present.
This diversity remains unexplained and larger samples are needed to study possible systematic trends or identify sub-groups.
A weak relation between X-ray and bolometric luminosity is present, but the 
scatter in X-ray brightness around $L_{\rm bol} \gtrsim 300~L_{\odot}$ is about two orders of magnitudes,
with the caveat that the threshold region has also the densest sampling due to selection strategies of the previous observing campaigns.
One of the major open questions is, which stellar parameters determine the X-ray luminosity of an Ap/Bp star.
MCWS models predict roughly a linear relation between X-ray luminosity and mass loss rate and mass loss rates depends in first order on stellar luminosity ($\dot{M} \propto L^{5/2}$).
The dependence on other parameters like the stellar radius, mass or magnetic field is moderate, differences up to factors of a few are expected in realistic cases.
Inspecting the dependence of $L_{\rm X}$ on other stellar properties (Fig.~\ref{figlx}, right panel), 
we find a very similar picture when replacing luminosity with related parameter like mass or effective temperature,
the corresponding threshold values are about 3.8~$M_{\odot}$ and 14000~K.

In contrast, stellar age or average magnetic field strength are at best of minor importance in setting the X-ray brightness in our sample stars. 
Trends that would relate X-ray brightness or even X-ray detections to an evolutionary phase are not present and already average magnetic fields of a few hundred Gauss are sufficient
to generate X-ray emission.
Also combinations of these parameters with e.g. luminosity do not show more convincing relations as the ones note above. 
The rotational data for our sample stars is less complete and where data exists no general trend that relates rotation to X-ray brightness is seen.
Remarkably, CU~Vir is the least luminous target with strong indications for X-ray emission and shows the fastest rotation, which might help to overcome gravity when launching the wind.

Given the limitations of the existing X-ray and stellar data and the current models, the large scatter between the individual targets remains mostly unexplained.
This might be indicative of multiple physical regimes for the magnetospheres among the studied Ap/Bp stars, 
but the physical properties of confined plasma and disk like structures in extreme magnetospheres have so far been studied only 
rudimentarily. If true, details of the magnetic field configuration and its obliquity must play a significant role in setting the 
individual X-ray properties.
In addition, the viewing geometry could also influence the observed X-ray characteristics and undiscovered binaries in the sample
can easily contaminate the observed X-ray brightness distribution and are a viable explanation, especially for the more extreme targets.

A related open question is the origin of the apparent threshold between the predominantly X-ray faint and the X-ray bright populations. 
Given the large scatter in X-ray brightness of the detected stars and the tight upper limits of several undetected stars,
attributing the repeated non-detection of virtually all stars below $250~L_{\odot}$ to
lacking sensitivity of the X-ray observations is a unlikely explanation.
In the classical MCWS model the X-ray luminosity follows the relation $L_{\rm X}\propto \dot M V_{\infty} B^{0.4}_{*}$, 
i.e. it is mainly linearly dependent on mass loss rate and wind speed. 
Further, the plasma temperature follows $T_{\rm X} \propto V_{\rm sh}^{2}$ and it needs to be sufficiently high enough to produce detectable X-ray emission.
Considering these correlations, a significant reduction in mass-loss rate or wind speed would be required to adequately affect the rotating magnetospheric structures and make the star sufficiently X-ray faint.

Theoretical mass loss models predict a strong decline in wind efficiency and mass loss rate at temperatures below 20~kK and at some point
the stellar wind has to become insufficient to generate X-ray emission, even in the presence of strong magnetic fields. \cite{bab95} derive 14 kK as the threshold for the onset of a non-separated wind, i.e. acceleration of all elements. 
This is very close to the threshold temperature that is seen in the X-ray data of Ap/Bp stars and a strong drop in mass-loss rate is expected to be connected to a strong decline in X-ray brightness.

Further, recent MCWS studies have shown that in the regime of weak winds, a so-called shock retreat might significantly reduce or
even quench the production of X-ray emission \citep{dou14}. The shock retreat works along the magnetic field lines and reduces the pre-shock velocities, 
resulting in fainter and softer X-ray emission. Shock retreat is especially effective at low mass loss rates, but the predicted changes are still quite gradual.
While it will definitely help to explain the observed trend, it might be especially
important in the fainting and softening of the X-ray emission towards less luminous stars within the detected objects.
In summary, several mechanisms have been proposed that result in a reduction of X-ray brightness at the observed threshold and
adding them together might cause a strong drop or even discontinuity in the X-ray detections.

\begin{figure}
\begin{center}
\includegraphics[width=84mm]{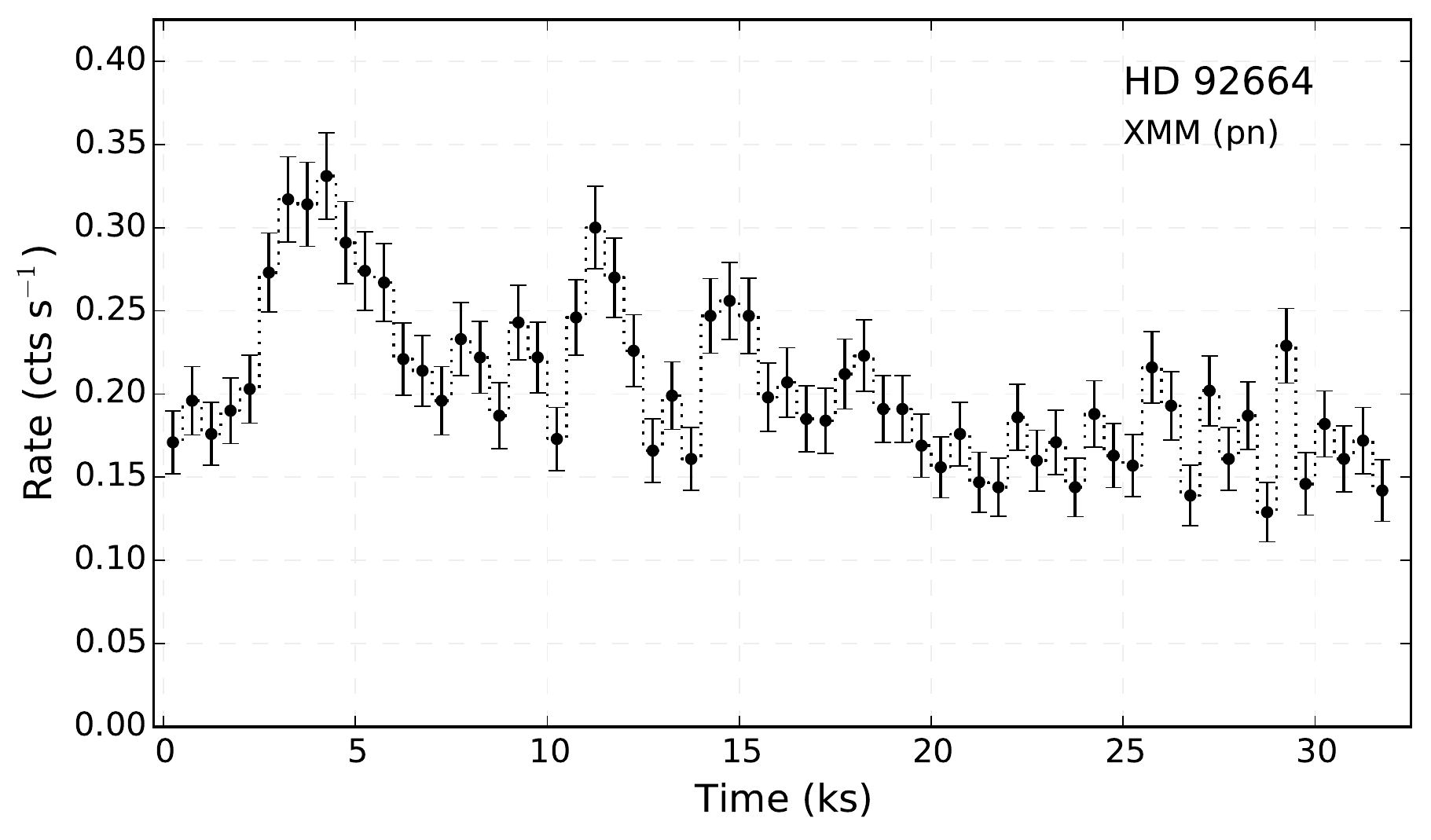}
\end{center}
\caption{\label{lc2}X-ray light curve of HD 92664 (V364 Car); XMM-Newton pn data with 500~s binning.}
\end{figure}

Variability studies at X-ray energies are biased towards the brighter targets and longer exposures, 
but if sufficient photons are detected most of our targets show at least minor X-ray variability of a few ten percent up to a factor of two on 
timescales of hours. In Fig.~\ref{lc2} we show an example from one of the newly observed stars. The light curves typically have shapes that resemble magnetic activity, 
however an event like the one seen on IQ~Aur is so far unique in our Ap/Bp star sample.

Gradual trends or rotational modulation are not seen in the X-ray data, but most observations are short compared to typical rotational timescales
of about one to a few days. The IQ~Aur observation covers more than half a rotation period without any clear sign of modulation, but viewing geometry and obliquity are uncertain and the large extend of the presumed X-ray emitting region could dilute rotational effects. Significantly extended observational campaigns would be required to more generally address these phenomena.

\begin{landscape}
\begin{table}
\caption{\label{results}X-ray properties and stellar data of Ap/Bp stars used in the analysis.}
\setlength\tabcolsep{4pt}{
\begin{tabular}{llllrrrrrrrrrrr}
\\\hline
& Star & Name & SpT & \multicolumn{1}{c}{d} & $\log T_{\rm eff}$ & $\log L_{\rm bol}$ & \multicolumn{1}{c}{$M$} & log\,t & \multicolumn{1}{c}{$<Be>$}  & $\log L_{\rm X}$ & $\log L_{\rm X}/L_{\rm bol}$& $F_{\rm X}$ & \multicolumn{1}{c}{$kT$} & counts\\
& &   & &[\,pc\,] &  [\,K\,]&  [\,$L_{\odot}$\,] & [\,$M_{\odot}$\,] & [\,yr\,] & [\,G\,] & [\,erg\,s$^{-1}$\,] & & [\,erg\,s$^{-1}$\,cm$^{-2}$\,]& [\,keV\,] & \\\hline
1 & HD 12767$^{*}$ &  $\nu$ For & B9.5/A0 Si&   110 &  4.11 &  2.47 &  3.75  & 8.13  &   242 & 29.5 & -6.5 &  2.4e-13  & 0.75  & 160\\
2 & HD 22470  & EG Eri & B9 Si   &  145 & 4.12  & 2.41 &  3.67  & 8.10  &   733   & $<$ 27.4 & $<$ -8.6& $<$ 1.0e-15  & - & - \\
3 & HD 25823$^{*}$ & GS Tau & B9 SrSi &  151&  4.11  & 2.50 &  3.80  & 8.13  &   668  &  28.9 & -7.2 & 3.0e-14  & 0.8& 90  \\
4 & HD 27309 & V724 Tau &A0 SiCr  &  96 &  4.08  & 2.01 &  3.04  & 8.07  &  1356 & $<$ 26.9 & $<$ -8.2 & $<$ 7.5e-16  & -& - \\
5 & HD 28843  &DZ Eri &B5/B9  Si He-w&     131 &  4.14  & 2.30  & 3.67  & 7.65  &   344   & $<$ 27.2 & $<$ -8.6 & $<$ 9.3e-16  & - & - \\
6 & HD 34452 & IQ Aur& A0/B9 Si &     127 &   4.16  & 2.49  & 4.02  & 7.82  &   743 &  29.6& -6.5 & 1.9e-13  & 0.9 & 14500 \\
7 & HD 40312  &$\theta$ Aur &A0 Si&      53 &  4.01  & 2.42  & 3.41  & 8.38  &   486   &$<$ 26.8 & $<$ -9.2 &  $<$ 1.8e-15  & -& -  \\
8 & HD 54118 & V386 Car  &A0 Si&      86 &  4.02  & 1.89 &  2.72 &  8.45 &   1020 & $<$ 27.2 &$<$ -8.2&  $<$  2.0e-15  & -& - \\
9 & HD 73340 & HV Vel &B8/B9 Si&     143 &   4.15 &  2.37  & 3.77 & 7.79  & 1644 & $<$ 29.0 &$<$ -6.9&  $<$  3.5e-14  & -& - \\
10& HD 92664  & V364 Car&B9 Si&      142 &   4.16  & 2.48 &  3.97  & 7.84  &   803   &30.1& -5.9&  5.4e-13  & 1.45 & 6500 \\
11& HD 105382$^{*}$ &V863 Cen &B6/B5 He-w&   115&    4.21 &  2.85 &  4.93 &  7.69  &   751 &  29.2& -7.2 &  1.0e-13 &  0.6 & 290 \\
12& HD 112413 &$\alpha^{2}$ CVn &A0p SiCrEu&      33 &   4.06  & 2.03 &  2.98  & 8.27  &  1448  &$<$ 26.0 &$<$ -9.6& $<$  7.4e-16 & - & - \\
13& HD 124224$^{*}$ & CU Vir&B8/B9 Si& 80 &  4.08  & 2.00 &   3.00 &   7.97 &    572 &  28.4& -7.2 &  3.0e-14 &  1.25& 280  \\
14& HD 125823 &  V761 Cen &B7III/B5 He-w&   128 &   4.25  & 3.02  & 5.57  & 7.51  &   469  & 28.8 & -7.8&  3.5e-14  & 0.6 & 70 \\
15& HD 133880 & HR Lup&A0/B9 Si&   126 &    4.08  & 2.07 &  3.10 &   8.15  &  2415  &$<$ 28.6 &$<$ -7.1& $<$ 2.1e-14  & - & - \\
16& HD 137509$^{*}$ & NN Aps&B9/B8 SiCrFe He-w & 249 &  4.15 &  2.44 &  3.88 &  7.89 &   1020 &  28.0& -8.0 &   1.4e-15 &  0.5 & $\lesssim 10$ \\
17& HD 142301 &V927 Sco &B8III/B8 Si He-w&   139 &  4.19  & 2.53 &  4.30  &  7.21  &  2104 & 28.1 & -8.0 &   5.5e-15 &  0.5 & 25 \\
18& HD 143473 &LL Lup &B9 Si&     123 &  4.11  & 1.82 &  3.00  &  6.44  &  4293  & $<$ 27.1& $<$ -8.3& $<$  7.5e-16  & - & - \\
19& HD 144334 &V929 Sco &B8 Si He-w&     149 &  4.17  & 2.50  &  4.08 &  7.72  &   783 & 30.4 & -5.7 &  9.0e-13  & 5.0& 500 \\
20& HD 146001 & -&B7IV/B8 He-w&   140&  4.15  & 2.41 &  3.90 &   6.70  &    647  & 28.9& -7.1 & 3.5e-14  & 1.0  & 25\\
21& HD 147010 & V933 Sco &B9 SiCrSr&     143&  4.12  & 2.04 &  3.23  & 6.47  &  4032  &$<$ 27.3 &$<$ -8.4&  $<$ 7.3e-16  & - & - \\
22& HD 175362 & V686 CrA &B4IV/B6 Si He-w&   130 &   4.22  & 2.64  &  4.64  & 6.92  &  3738  & 28.2 & -8.0&  8.0e-15 &  0.8& 20 \\
23& HD 184905 & V1264 Cyg &A0 SiCr&     165  & 4.04  & 1.93 &  2.80  & 8.39 &  5052  &$<$ 27.4 &$<$ -8.1& $<$ 7.3e-16  & - & - \\
24& HD 217833 & V638 Cas&B9III/B8 SiCr He-w&  221  & 4.17 & 2.59 &  4.23 & 7.83 &   3650 &  $<$ 28.4 &$<$ -7.8& $<$ 3.0e-15  & - & - \\\hline
\end{tabular}}
$^{*}$ see notes. Stellar data is mainly taken from \cite{koch06}, magnetic field data from \cite{by09}, see text for details.
\end{table}
\end{landscape}

\subsection{X-ray spectra of Ap/Bp stars}

Spectral information allows to further constrain the nature of the detected X-ray emission.
Unfortunately, only for a few of the detected targets the quality of the X-ray data allows
a detailed spectral modeling and with the exception of IQ~Aur, high-resolution X-ray grating spectra are not available at sufficient quality.
Therefore the analysis presented here uses the medium resolution spectra obtained from the EPIC/pn (XMM) and ACIS (Chandra) detector.

It is worth recalling that the initial motivation for the MCWS model was the detection of IQ~Aur as an X-ray bright target with constant and very soft emission,
dominated by a cool 3.5~MK plasma with $T_{\rm X} = 0.3$~keV \citep{bab97}. 
A strong cool component is also seen in the XMM-Newton data of IQ~Aur, but in addition significant amounts of hotter plasma ($T_{\rm X} \gtrsim 1.0$~keV) 
are detected in quasi-quiescence, i.e. outside the large flare \citep{rob11}.
With respect to plasma temperature, IQ~Aur is not unique or an extreme case in our stellar sample. For example, the B9p star HD~92264 has nearly identical stellar parameters, but it is 
three times X-ray brighter and dominated by even hotter plasma.

\begin{figure}[t]
\begin{center}
\includegraphics[trim={0cm 1.2cm 0cm 0cm}, clip,width=50mm, angle=-90]{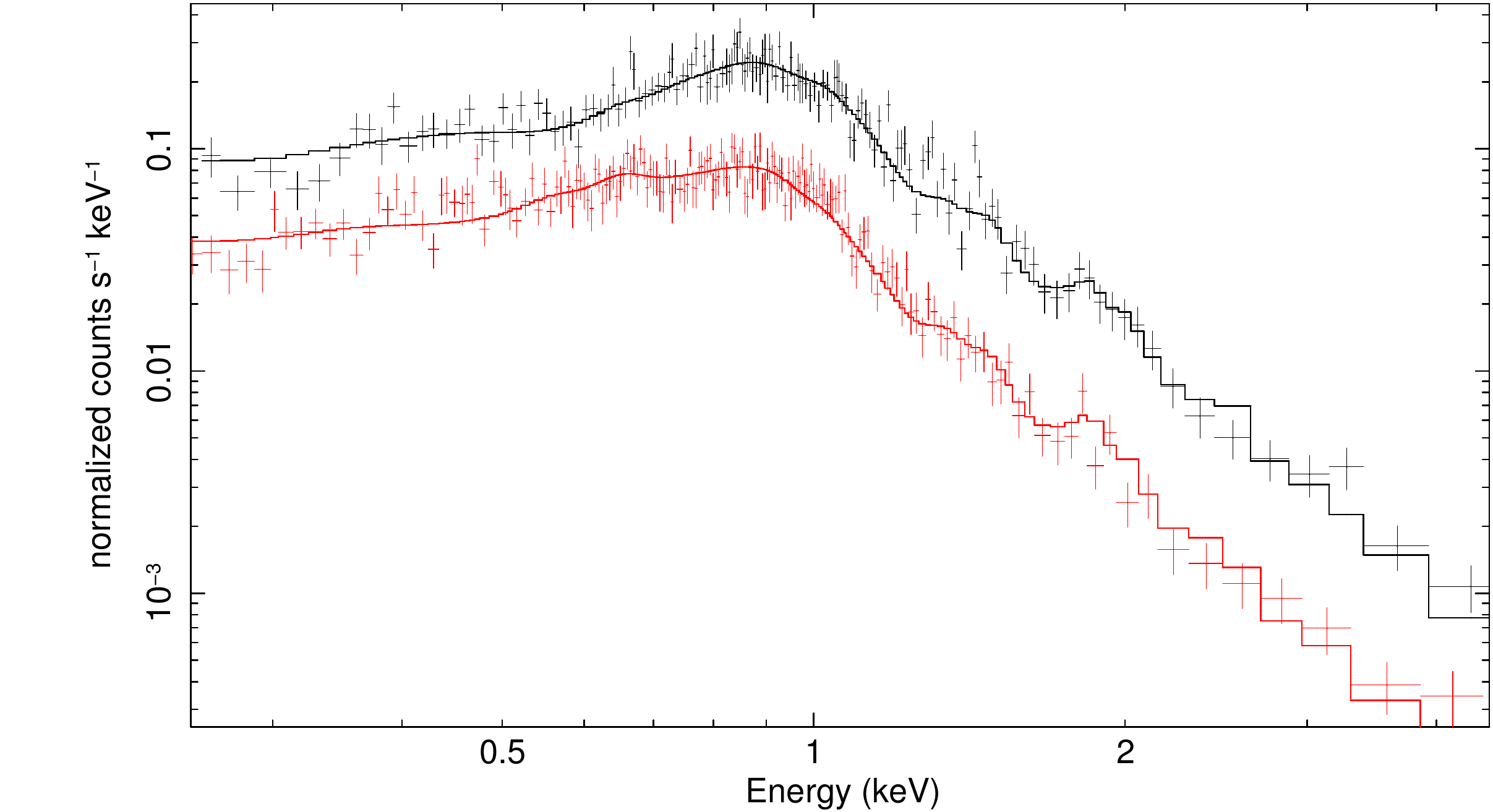}
\includegraphics[trim={0cm 1.2cm 0cm 0cm}, clip,width=50mm, angle=-90]{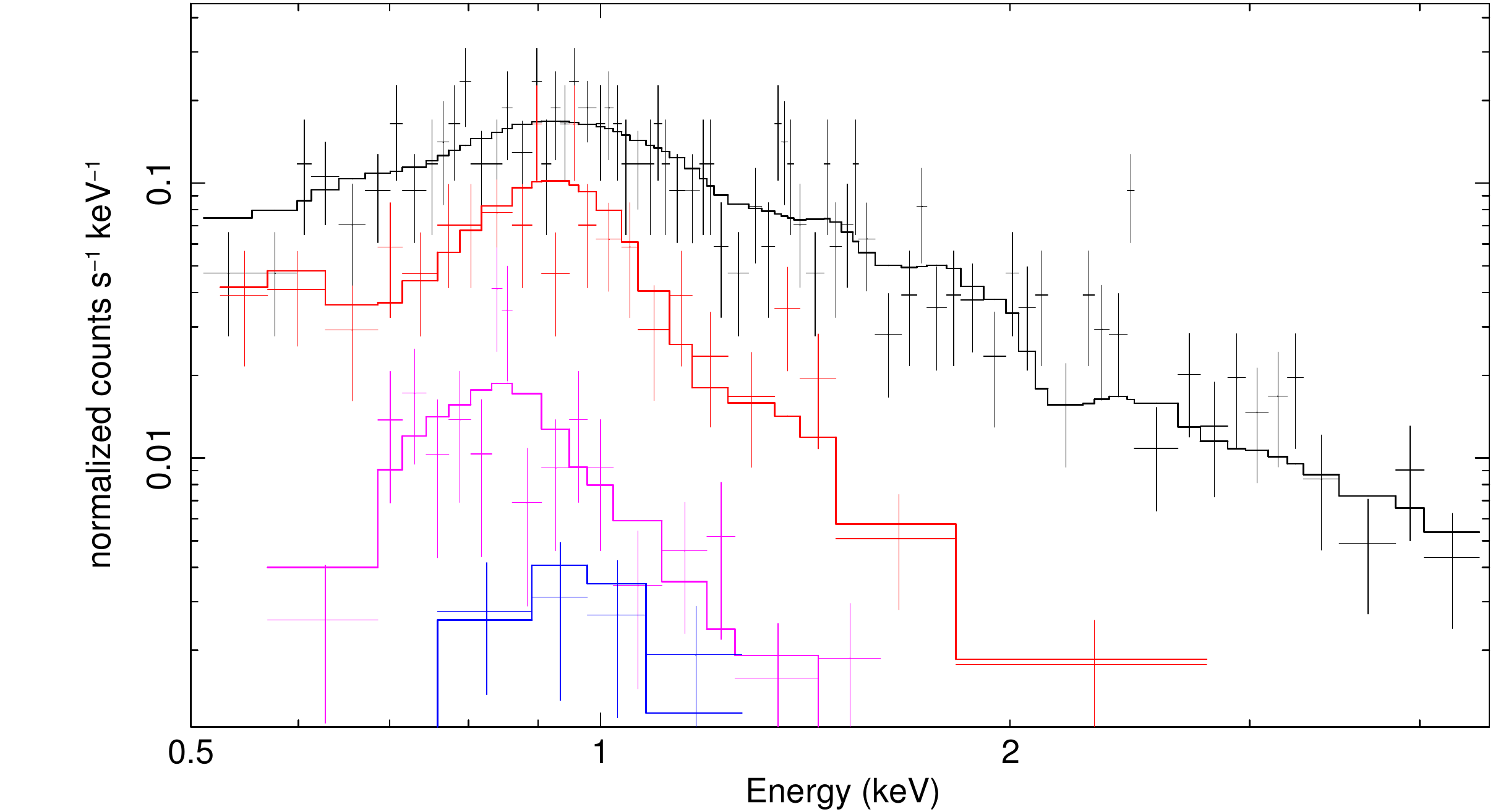}
\end{center}
\caption{\label{figspec}X-ray spectra of Ap/Bp stars. {\it Top:} XMM pn spectra of HD~92264/V364~Car (black) and HD 34452/IQ~Aur (red, quasi-quiescence).
{\it Bottom:} Chandra ACIS spectra of HD~144334/V929~Sco (black), HD~12767/$\nu$~For (red), HD~125823/V761 Cen (magenta) and HD~175362/V686~CrA (blue).}
\end{figure}

A comparison of X-ray spectra from Ap/Bp stars is shown in Fig.~\ref{figspec}. 
In the top panel we show XMM/pn spectra obtained during the quasi-quiescent state of IQ~Aur and the one of V364~Car. 
Both stars have similar plasma temperatures in multi-temperature models, but with very different distribution. Using a 2-T model, we find that
the cool plasma component ($\sim 3$~MK) is virtually identical, but the intermediate (8\,--\,10~MK) and especially the hot (15\,--\,20~MK) component are more pronounced
in the X-ray brighter star. The rough emission measure ratios in the respective spectral models are 1:1:1 for IQ~Aur and 1:3:7 for V364~Car. 

In the lower panel we show four Chandra/ACIS spectra from Ap/Bp stars with a two orders of magnitude spread in X-ray luminosity. 
By far the brightest target in our sample, V929~Sco, exhibits also the hardest spectrum with $T_{\rm X} \sim 5$~keV. 
Temperatures of around 50~MK are out of reach for wind shocks for late B stars, even assuming head-on collision at terminal wind speed. 
The spectrum of V929~Sco resembles the flare spectrum of IQ~Aur, but remarkably the light curve shows no significant variability that would be indicative of an ongoing strong flare during the short 3~ks observation. These extreme spectral properties require that the magnetospheres of a late Bp stars can generate significant magnetic activity or that these objects are indeed multiple systems that contain highly active young solar-type stars.

\begin{figure}
\begin{center}
\includegraphics[width=84mm]{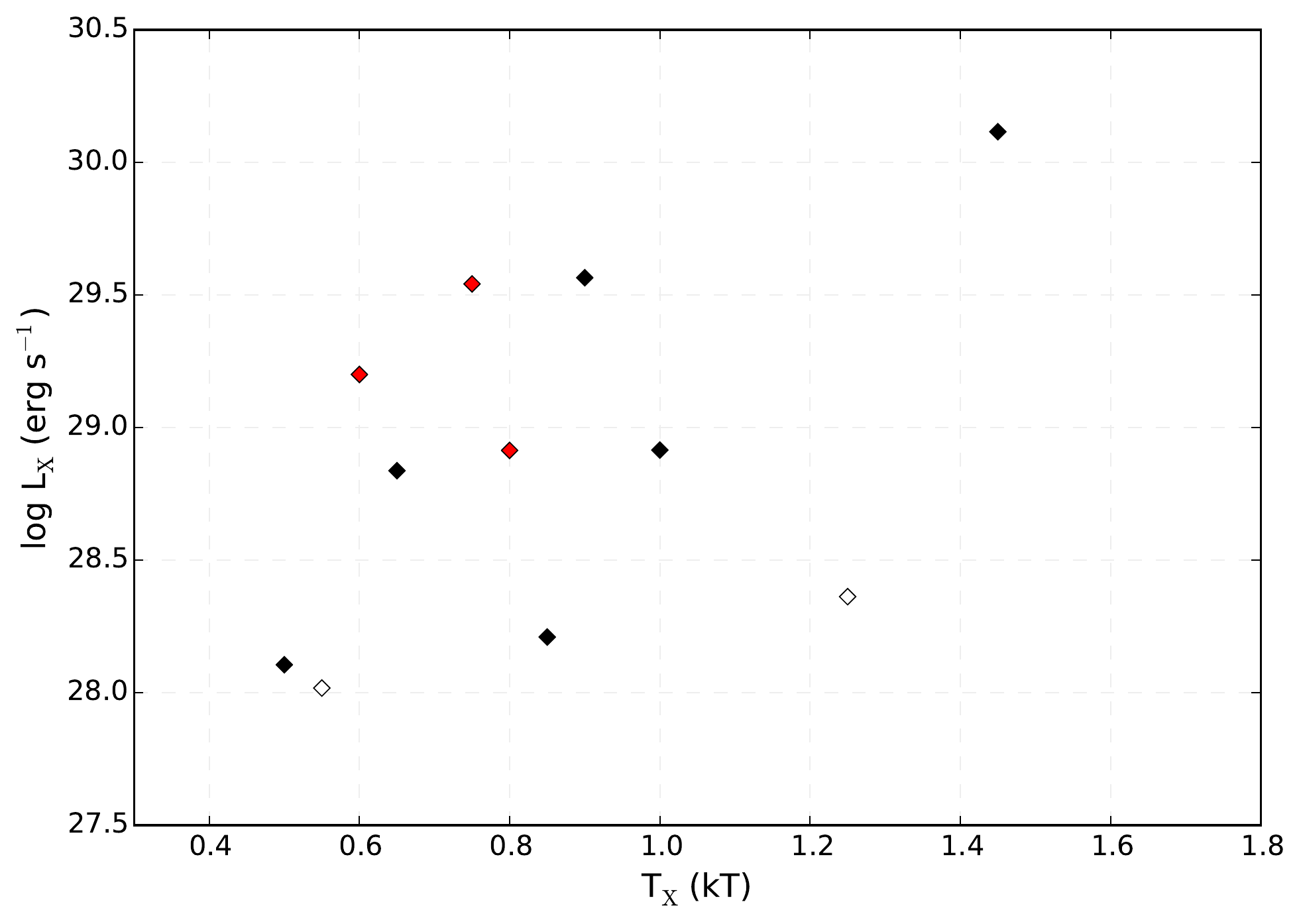}
\end{center}
\caption{\label{kt}X-ray brightness vs. spectral hardness of the detected stars, symbols are identical to Fig.\,\ref{figlx}. HD~144334/V929~Sco with $kT =5$~keV is outside the plotting range.}
\end{figure}

Inspecting the average plasma temperatures of the detected sources, we find that X-ray brightness and spectral hardness are only moderately correlated, see Fig.~\ref{kt}.
The softer spectra are found mostly in the fainter targets of our sample, but errors become quite large for targets with low counts.
The typical average plasma temperatures for the detected Ap/Bp stars are in the range of 5\,--\,10~MK
and none of our targets shows an average plasma temperature that is significantly below 5~MK.

Absorption and metallicity/elemental abundances are the remaining potentially free parameters in our spectral models.
X-ray absorption is typically low to moderate in the studied targets, we therefore neglect it in our analysis. 
If included in spectral modeling, derived column densities are typically $N_{\rm H} \approx 0 \dots 1 \times 10^{20}$~cm$^{-2}$, consistent with absorption from interstellar matter.
The effects are overall minor, when adding an absorption component would slightly increase the intrinsic X-ray luminosity and lower the average X-ray temperature in all cases.
Results become arbitrary for very low-quality data and stronger effects for these targets cannot be ruled out completely, but are considered unlikely. 
The abundances of individual elements within the X-ray emitting plasma cannot be constrained for almost all of our targets 
and even the determination of a global metallicity is mostly ambiguous, therefore we adopt solar values for our targets.
If left as a free parameter, we find values around 0.4~solar metallicity for IQ~Aur and V364~Car from their medium resolution spectra.
However, a reduction in metallicity is largely compensated by an increase in emission measure and the effects on average temperature and X-ray luminosity are again minor.

Overall, the medium resolution X-ray spectra of Ap/Bp stars are quite similar to those of low-mass stars. Significantly
larger observational efforts and high resolution X-ray spectra are required to reveal the more subtle spectral differences that would allow to distinguish between the
different physical processes involved in generating X-ray emission.

\subsection{The X-ray Radio connection}

The presence of stellar magnetospheres provides not only an explanation for the X-ray but also for the bright radio emission of Ap/Bp stars \citep{lin92}.
The radio emission mechanism in Ap/Bp stars is mainly gyrosynchrotron as can be expected from shocked winds where Fermi-acceleration produces fast electrons
instead of or in addition to those from magnetic reconnections. The radio emission was found be be variable up to a factor of a few.
Among the Ap/Bp stars surveyed with the VLA are several of the here presented objects. A comparison shows
that predominantly the hotter Ap/Bp stars are detected as bright radio sources and that typically the radio brighter stars are seen as X-ray sources.

This is quantitatively similar to the X-ray data, albeit no direct correlation between X-ray and radio brightness is found.
Radio emission correlates overall better with stellar wind parameters, the present scatter is about one magnitude in the studied sample.
Further, compared to coronal sources Ap/Bp stars strongly violate the X-ray/Radio connection, also known as Guedel-Benz-Relation \citep{gued93}. 
Magnetic activity produces, within a factor of a few, a tight correlation between X-ray and radio brightness that is valid over many orders of magnitude.
Compared to coronal sources Ap/Bp stars are typically found to be radio overluminous by factor of ten up to hundred.
Faint radio detections are reported by \citep{dra06} also for several of the X-ray undetected stars in our sample.
This was interpreted as evidence for the companion hypothesis, specifically based on the X-ray non-detection of HD~27309 and HD~133880.
However, it might also indicate the presence of a different or additional emission mechanisms that predominantly creates radio emission.
Both radio and X-ray emission would then be intrinsic and originate in the complex disk-harboring magnetospheres of Ap/Bp stars.

Rotationally modulated or even pulsed radio emission is seen in several Ap/Bp stars. The prominent example is  
the Ap star CU~Vir, it show clear signs for auroral radio emission that has been interpreted as electron-cyclotron maser emission.
It is an oblique rotator and due to the 'lighthouse effect' radio pulses are seen from a highly collimated beam of coherent, 100\% polarized radio emission \citep{tri11}. 
Its strong magnetic field and extremely fast rotation make CU~Vir a perfect test-bed for several fundamental astrophysical processes. 
If its X-ray detection can be confirmed, it would be the star that is furthest away from the transition line.

\section{Outlook}

Further X-ray observations of Ap/Bp stars are initiated to help answering some of the open questions. However, only a few targets are typically observed in pointed observations and
the expected increase in sample size by this approach is limited. Nevertheless, observations of key targets can deliver deeper insights into the X-ray generation mechanisms at work.
In the multi-wavelength perspective, spectropolarimetric campaigns and radio observations could help to detect and characterize magnetic stars. 
While current studies often concentrate on massive stars, an extension deeper into the intermediate mass range would be a valuable step for future efforts. 
In addition, searches for possible companions in these systems via astrometric, interferometric or spectroscopic means are desired to check for an external contribution or origin of the detected X-ray emission. 

With respect to X-ray coverage the future X-ray all-sky survey by eROSITA \citep{mer12} is a promising alternative. A more sensitive X-ray survey
will provide a more complete and much deeper coverage of the stellar Ap/Bp parameter space with X-ray observations and identify new targets that warrant a detailed investigation with pointed follow-up observations. With the expected sensitivity of the eROSITA all-sky survey, about 70\,\% of the X-ray sources presented here would have been detected.
When considering spectroscopic studies of Ap/Bp stars,
the sensitivity of the current X-ray instrumentation at its limit and very deep exposures are required to obtain meaningful data. Alternatively one has to await more sensitive missions like Athena, that are planned for the next decade and will open new prospects in X-ray astronomy. Possible observational tests include the search for rotational modulation of the X-ray emission, the determination of more precise elemental abundances and emission measure distributions of the X-ray emitting plasma.

From the theoretical and computational side, the recent developments in modeling magnetospheres are promising in characterizing X-ray emission within the MCWS, but full MHD (or even 3D-MHD) simulations are currently computationally out of reach for the highly magnetic objects like IQ~Aur.
Here especially the rigid-field hydrodynamic approach is a promising alternative to study the regime of Ap/Bp stars, where typically the dominance of the magnetic field over the wind is at its extreme.
Future RFHD simulation could enable a much deeper study of the large, strong field magnetospheres including rotation or complex field geometries
to shed light on the complex phenomena observed in intermediate mass magnetic stars.

\bigskip
\noindent
{\it Acknowledgements. }{\small JR acknowledges support from DLR under 50QR0803. I thank H.M.~G{\"u}nther, A. Dupree and E.~Adams for organizing, obtaining and analyzing the MMT observation of IQ~Aur. This research has made use of the VizieR catalogue access tool, CDS, Strasbourg, France.}

%
%



%


\section*{References}

\begin{thebibliography}{}

\bibitem[Abt \& Snowden(1973)]{abt73}
Abt, H.A. \& Snowden, M.S,
The Binary Frequency for AP Stars,
ApJS, 25, 137-162, 1973.

\bibitem[Alecian et al.(2013)]{ale13}
Alecian, E., Wade, G.A., Catala, C., et al,
A high-resolution spectropolarimetric survey of Herbig Ae/Be stars - I. Observations and measurements,
MNRAS, 429, 1001-1026, 2013.

\bibitem[Alecian et al.(2014)]{ale14}
Alecian, E., Kochukhov, O., Petit, V. et al,
Discovery of new magnetic early-B stars within the MiMeS HARPSpol survey,
A\&A, 567, A28 (19p), 2014.

\bibitem[Babcock(1960)]{bab60}
Babcock, H.W., 
The 34-KILOGAUSS Magnetic Field of HD 215441,
ApJ, 131, 521-532, 1960.

\bibitem[Babel(1995)]{bab95}
Babel, J., 
Multi-component radiatively driven winds from A and B stars. I. The metallic wind of a main sequence A star
A\&A, 301, 823+, 1995.

\bibitem[Babel \& Montmerle(1997)]{bab97}
Babel, J. \& Montmerle, T. 1997, 
X-ray emission from Ap-Bp stars: a magnetically confined wind-shock model for IQ Aur,
A\&A, 323, 121-138, 1997.

\bibitem[Bychkov et al.(2009)]{by09}
Bychkov, V.D., Bychkova, L.V. \& Madej, J.,
Catalogue of averaged stellar effective magnetic fields - II. Re-discussion of chemically peculiar A and B stars,
MNRAS, 394, 631-642, 2009.

\bibitem[Chini et al.(2012)]{chi12}
Chini, R., Hoffmeister, V.H., Nasseri, A. et al.,
A spectroscopic survey on the multiplicity of high-mass stars,
MNRAS, 424, 1925-1929, 2012.

\bibitem[Czesla \& Schmitt(2007)]{cze07}
Czesla, S. \& Schmitt, J.H.M.M.,
Are magnetic hot stars intrinsic X-ray sources?,
A\&A, 465, 493-499, 2007.


\bibitem[Drake et al.(1994)]{dra94}
Drake, S.A., Linsky, J.L., Schmitt, J.H.M.M. \& Rosso, C.,
X-ray emission from chemically peculiar stars,
ApJ, 420, 387-391, 1994.


\bibitem[Drake et al.(2006)]{dra06}
Drake, S.A., Wade, G.~A. \& Linsky, J.L.,
The Radio and X-ray Properties of Magnetic Bp/Ap Stars,
Proc. X-ray Universe 2005, ESASP 604, 73-74, 2006.

\bibitem[Freyhammer et al.(2008)]{fre08}
Freyhammer, L.M., Elkin, V.G., Kurtz, D.W, et al.,
Discovery of 17 new sharp-lined Ap stars with magnetically resolved lines
MNRAS, 389, 441-460, 2008.

\bibitem[Gagn{\'e} et al.(2005)]{gag05}
Gagn{\'e}, M., Oksala, M.E., Cohen, D.H, et al.,
Chandra HETGS Multiphase Spectroscopy of the Young Magnetic O Star ${\theta}^{1}$ Orionis C
ApJ, 628, 986-1005, 2005.

\bibitem[Garc{\'e}s et al.(2011)]{gar11}
Garc{\'e}s, A., Catal{\'a}n, S. \& Ribas, I.,
Time evolution of high-energy emissions of low-mass stars
A\&A, 531, A7 (10p), 2011.


\bibitem[G{\"u}del \& Benz(1993)]{gued93}
G{\"u}del, M. \& Benz, A.O,
X-ray/microwave relation of different types of active stars.
ApJL, 405, L63-L66, 1993.

\bibitem[G{\"u}del \& Naz{\'e}(2009)]{gued09}
G{\"u}del, M. \& Naz{\'e}(2009),
X-ray spectroscopy of stars,
A\&AR, 17, 309-408, 2009.

\bibitem[Kochukhov(2006)]{koch06a}
Kochukhov, O.,
Remarkable non-dipolar magnetic field of the Bp star HD 137509,
A\&A, 454, 321-325, 2006.

\bibitem[Kochukhov \& Bagnulo(2006)]{koch06}
Kochukhov, O. \& Bagnulo, S.,
Evolutionary state of magnetic chemically peculiar stars,
A\&A, 450, 763-775, 2006.

\bibitem[Landstreet(1992)]{lan92}
Landstreet, J.D.,
Magnetic fields at the surfaces of stars,
A\&AR, 4, 35-77, 1992.

\bibitem[Landstreet et al.(2007)]{lan07}
Landstreet, J.D., Bagnulo, S., Andretta, V., et al.,
Searching for links between magnetic fields and stellar evolution: II. The evolution of magnetic fields as revealed by observations of Ap stars in open clusters and associations,
A\&A, 470, 685-698, 2007.

\bibitem[Linsky et al.(1992)]{lin92}
Linsky, J.L., Drake, S.A. \& Bastian, T.S,
Radio emission from chemically peculiar stars,
ApJ, 393, 341-356, 1992.

\bibitem[Mason et al.(2001)]{mas01}
Mason, B.D., Wycoff, G.L., Hartkopf, W.I, et al.,
The Washington Double Star Catalog,
AJ, 122, 3466-3471, 2001.

\bibitem[Merloni et al.(2012)]{mer12}
Merloni, A., Predehl, P., Becker, W., et al.,
eROSITA Science Book: Mapping the Structure of the Energetic Universe,
ArXiv e-prints, 1209.3114, 2012.

\bibitem[Mullan(2009)]{mul09}
Mullan, D.J.,
Flares on a Bp Star,
ApJ, 702, 759-766, 2009.

\bibitem[Naz{\'e} et al.(2014)]{naze14}
Naz{\'e}, Y., Petit, V., Rinbrand, M., et al.,
X-Ray Emission from Magnetic Massive Stars,
ApJS, 215, 10 (20p), 2014.

\bibitem[Oskinova et al.(2011)]{osk11}
Oskinova, L.M., Todt, H., Ignace, R., et al.,
Early magnetic B-type stars: X-ray emission and wind properties,
MNRAS, 416, 1456-1474, 2011.


\bibitem[Petit et al.(2013)]{pet13}
Petit, V., Owocki, S.P., Wade, G.A., et al.,
A magnetic confinement versus rotation classification of massive-star magnetospheres,
MNRAS, 429, 398-422, 2013.


\bibitem[Pourbaix et al.(2004)]{pou04}
Pourbaix, D., Tokovinin, A.A., Batten, A.H, et al.,
SB9: The ninth catalogue of spectroscopic binary orbits,
A\&A, 424, 727-732, 2004.

\bibitem[Renson \& Manfroid(2009)]{ren09}
Renson, P. \& Manfroid, J.,
Catalogue of Ap, HgMn and Am stars,
A\&A, 498, 961-966.


\bibitem[Robrade \& Schmitt(2009)]{rob09}
Robrade, J. \& Schmitt, J.H.M.M.,
Altair - the 'hottest' magnetically active star in X-rays,
A\&A, 497, 511-520, 2009.

\bibitem[Robrade \& Schmitt(2011)]{rob11}
Robrade, J. \& Schmitt, J.H.M.M.,
New X-ray observations of IQ Aurigae and $\alpha^{2}$ Canum Venaticorum. Probing the magnetically channeled wind shock model in A0p stars,
A\&A, 531, A58 (11p), 2011.


\bibitem[Sanz-Forcada et al.(2004)]{san04}
Sanz-Forcada, J., Franciosini, E. \& Pallavicini, R.,
XMM-Newton observations of the {$\sigma$} Ori cluster.
A\&A, 421, 715-727, 2004.

\bibitem[Stelzer et al.(2006)]{ste06}
Stelzer, B., Hu{\'e}lamo, N., Micela, G. \& Hubrig, S.,
Testing the companion hypothesis for the origin of the X-ray emission from intermediate-mass main-sequence stars,
A\&A, 452, 1001-1010, 2006.


\bibitem[Townsend et al.(2007)]{tow07}
Townsend, R.H.D, Owocki, S.P. \& ud-Doula, A., 
A Rigid-Field Hydrodynamics approach to modelling the magnetospheres of massive star,
MNRAS, 382, 139-157, 2007.

\bibitem[Trigilio et al.(2011)]{tri11}
Trigilio, C., Leto, P., Umana, G. et al.,
Auroral Radio Emission from Stars: The Case of CU Virginis,
ApJL, 739, L10 (5p), 2011.

\bibitem[ud-Doula \& Owocki(2002)]{dou02}
ud-Doula, A. \& Owocki, S.P.,
Dynamical Simulations of Magnetically Channeled Line-driven Stellar Winds. I. Isothermal, Nonrotating, Radially Driven Flow,
ApJ, 576, 413-428, 2002.

\bibitem[ud-Doula et al.(2006)]{dou06}
ud-Doula, A., Townsend, R.H.D \& Owocki, S.P.,
Centrifugal Breakout of Magnetically Confined Line-driven Stellar Winds
ApJL, 640, L191-L194, 2006.

\bibitem[ud-Doula et al.(2008)]{dou08}
ud-Doula, A., Owocki, S.P. \& Townsend, R.H.D,
Dynamical simulations of magnetically channelled line-driven stellar winds - II. The effects of field-aligned rotation,
MNRAS, 385, 97-108, 2008.

\bibitem[ud-Doula et al.(2014)]{dou14}
ud-Doula, A., Owocki, S.P., Townsend, R.H.D \& Petite, V.,
X-rays from magnetically confined wind shocks: effect of cooling-regulated shock retreat,
MNRAS, 441, 3600-3614, 2014.


\end{thebibliography}
\end{document}